\documentstyle[12pt,preprint,tighten,floats,aps,epsf,psfig]{revtex}

\tighten
\begin{document}
\draft
\title{Two-Higgs doublet models from TeV-scale supersymmetric extra U(1) 
models}
\author{D. A. Demir\footnote{Address after 1st of October:
ICTP, Trieste, Italy.}}
\address{Middle East Technical University, Department of Physics,
06531, Ankara, Turkey}
\date{\today}
\maketitle
\begin{abstract}
We investigate the reduction of a general TeV-scale supersymmetric extra 
$U(1)$ model to a 2HDM below the TeV- scale  through the tree level 
non-decoupling. Portions of the parameter space of the extra 
$U(1)$ model appropriate for obtaining a 2HDM are identified. Various 
properties of the resulting 2HDM are connected to the parameter space of 
the underlying model.\\
PACS: 12.60.Jv, 12.60.Cn, 12.60.Fr
\end{abstract}
\newpage
\section{Introduction}
For various theoretical and phenomenological reasons various extensions of 
the SM have been introduced. The simplest such extension is Two-Higgs 
Doublet Models (2HDM) where one extends the scalar sector of the SM by 
adding a second doublet without modifying the gauge structure 
\cite{haber}. Such models are mainly motivated by MSSM \cite{nilles} in 
which one has to introduce at least two Higgs doublets. One basic 
use of the 2HDM's is that one is able to create the necessary mass 
splitting between the  up- and down-type quarks \cite{froggat}. In this 
respect, various high energy processes have been discussed in detail 
\cite{kane}. Furthermore, the famous Sakharov conditions \cite{sakharov} 
for creating the baryon aymmetry in the universe cannot be satified in SM 
due to the smallness of the CP-violation. In 2HDM, however, explicit or 
spontaneous breaking of CP is allowed with the self interactions of the 
Higgs doublets \cite{kaplan}.
 
MSSM has two Higgs doublets and its Higgs sector mimics that of  
the 2HDM in many respects, except for the fact that supersymmetry fixes 
various parameters in terms of the gauge couplings \cite{haber}. However, 
MSSM has an hierarchy problem, namely, the scale of the supersymmetric 
mass term $\mu$ is not known. 

In this work, we work out supersymmetric models with an additional $U(1)$ 
which are known to solve the MSSM $\mu$ problem 
\cite{suematsu,lang,jose,dur}. Indeed, as was already argued in 
\cite{cvetic-lang}, in a large class of string models, breaking scale of 
the extra $U(1)$'s come out to be around a TeV. Whatever the Planck scale 
consideration (SUSY GUT's or Superstrings) from which these models 
follow, to be able to get a sensible solution to the problems mentioned 
above it is necessary to keep supersymmetry and the gauge symmetry exact 
till the TeV scale. When the SUSY is broken around the TeV scale one 
gets $\mu\sim {\cal{O}}$(TeV) so that the $\mu$-problem is avoided.
The possible existence of a supersymmetric Abelian gauge factor which 
is broken at the TeV scale together with the supersymmetry affects the 
weak scale observables through the non-decoupling effects in the Higgs 
sector \cite{ma-group}.  

In this work we analyze the Higgs sector of an Abelian extended SUSY 
model (we call this $Z^{\prime}$ model from now on) such that after the 
breaking of the extra supersymmetric $U(1)$ factor there arises an 
effective 2HDM at the weak scale. This problem has already been worked 
out in \cite{ma-group} where various phenomenological implications of 
non-decoupling effects in the Higgs sector were discussed. However, in this 
work we present a detailed analysis of the tree level constraints on the 
scalar potential of a general $Z^{\prime}$ model to be able to get a 2HDM 
below the TeV scale.

In Sec. 2 we first discuss the reduction of the scalar 
potential of a general $Z^{\prime}$ model to a 2HDM potential below the 
TeV scale. Next, we identify the appropriate portions of the $Z^{\prime}$ 
model parameter space to make this reduction process viable.

In Sec. 3 we discuss the properties of the resulting 2HDM potential in 
connection with several phenomenological issues.

In Sec. 4 we conclude the work.
\section{2HDM from a TeV scale supersymmetric extra $U(1)$}
In $Z^{\prime}$ models, the MSSM gauge group is extended to 
$G=SU(3)_{c}\times SU(2) \times U(1)_{Y} \times U(1)_{Y'}$ with the 
coupling constants $g_{3}$, $g_{2}$, $g_{Y}$, $g_{Y'}$, respectively.
Under $G$, the Higgs superfields are assigned the quantum numbers 
$\hat{H}_{1} \sim (1, 2, -1/2, {Q'}_{1})$, $\hat{H}_{2} \sim (1, 2, 1/2, 
{Q'}_{2})$, $\hat{S}\sim (1, 1, 0, {Q'}_{S})$. Here $\hat{S}$ is an SM 
gauge singlet whose vacuum expectation value (VEV) breaks the extra $U(1)$.
Part of the superpotential containing exclusively the Higgs fields is 
given by 
\begin{eqnarray}
W \ni h_{S}\,\hat{S}\,\hat{H}_{1}\cdot\hat{H}_{2}
\end{eqnarray}
In addition to this, the complete superpotential contains fermion trilinear 
mass terms. We discard such terms from the analysis because whenever the 
sfermions develop non-vanishing VEV's color and/or charge symmetries are 
broken. As long as the parameter spaces we work in do not imply 
non-vanishing sfermion VEV's, analysis of the Higgs dependent part of the 
potential suffices. We further note that due to the $U(1)_{Y'}$ symmetry 
a bare $\mu$ term is forbidden; thus form of the superpotential in (1) is 
unique. 
 
The full scalar potential in the $Z^{\prime}$ models is given by 
\begin{eqnarray}
V&=&{m_{1}^{0}}^{2}\mid H_{1}\mid ^{2}
+{m_{2}^{0}}^{2}\mid H_{2}\mid ^{2} + {m_{S}^{0}}^{2}\mid S\mid
^{2}+ \lambda^{0}_{1}\mid H_{1}\mid ^{4} +\lambda^{0}_{2}\mid
H_{2}\mid ^{4}\nonumber\\&+&\lambda^{0}_{S}\mid S\mid ^{4}+
\lambda^{0}_{12} \mid H_{1}\mid ^{2} \mid H_{2}\mid ^{2} +
\lambda^{0}_{1S} \mid H_{1}\mid ^{2} \mid S\mid ^{2} +
\lambda^{0}_{2S} \mid H_{2}\mid ^{2} \mid S\mid^{2}\nonumber\\&+&  
\tilde{\lambda}^{0}_{12}\mid H_{1}\cdot H_{2}\mid ^{2}-h_{s}A_{s}^{0}(S 
H_{1}.H_{2} + h. c.) 
\end{eqnarray}
where we discarded the hat on the superfields to denote their scalar 
components. In (2), ${m_{1}^{0}}^{2}$, ${m_{2}^{0}}^{2}$ and 
${m_{S}^{0}}^{2}$ are the soft  mass-squareds of $H_1$, $H_2$ and $S$, 
respectively.  $A_{s}^{0}$ is the Higgs trilinear cupling. While these 
quantities of mass dimension come from the soft supersymmetry breaking 
part of the potential, those terms involving the adimensional parameters 
$\lambda_{i}^{0}$ come from the supersymmetric part of the Lagrangian 
consisting of $F$ and $D$ terms: 
\begin{eqnarray}
\lambda_{1}^{0}&=&\frac{1}{8}G^{2}
+\frac{1}{2}{g_{Y'}}^{2}{Q'}_{1}^{2}\nonumber\\
\lambda_{2}^{0}&=&\frac{1}{8}G^{2}
+\frac{1}{2}{g_{Y'}}^{2}{Q'}_{2}^{2}\nonumber\\
\lambda_{S}^{0}&=&\frac{1}{2}{g_{Y'}}^{2}{Q'}_{2}^{2}\\
\lambda_{12}^{0}&=&\frac{1}{4}(g_{2}^{2}-g_{Y}^{2})
+{g_{Y'}}^{2}{Q'}_{1}{Q'}_{2}\nonumber\\
\lambda_{1S}^{0}&=&{g_{Y'}}^{2}{Q'}_{1}{Q'}_{S}+
{h_{s}}^{2}\nonumber\\
\lambda_{2S}^{0}&=&{g_{Y'}}^{2}{Q'}_{2}{Q'}_{S}+
{h_{s}}^{2}\nonumber\\
\tilde{\lambda}^{0}_{12}&=&h_{s}^{2}-g_{2}^{2}/2\nonumber
\end{eqnarray}
where $G=\sqrt{{g_{2}}^{2}+{g_{Y}}^{2}}$.
 
In this work, we shall not investigate the CP-violating or charge breaking 
properties of the vacuum state so we allow only the CP-even components of 
the Higgs fields to develop non-vanishing VEV's. Thus, VEV's of the Higgs 
fields in (2) are subjected to the following parametrization
\begin{eqnarray}
<H_{1}>=\frac{1}{\sqrt{2}}\left(\begin{array}{c c}
v_{1}\\0\end{array}\right)\; , \;
<H_{2}>=\frac{1}{\sqrt{2}}\left(\begin{array}{c c}
0\\v_{2}\end{array}\right)\; , \;
<S>=\frac{v_{s}}{\sqrt{2}}
\end{eqnarray}
with real $v_{1}, v_{2}$ and $v_{s}$. Furthermore by appropriate choice 
of the signs of the potential parameters they can always be taken 
positive \cite{lang}.

To be able to obtain an effective 2HDM below the TeV-scale, one should 
realize $v_s\sim {\cal{O}}$(TeV) and $v_{1}=0=v_{2}$ so that below 
the TeV scale SM- gauge group remains unbroken. Before an attempt to 
determine the portion of the parameter space for which ony $U(1){Y'}$ is 
broken, let us assume such a splitting between the SM-singlet and the 
doublet VEV's indeed exist and derive the effective 2HDM potential below 
the TeV-scale. For convenience we apply the parametrization
\begin{eqnarray}
S=\frac{1}{\sqrt{2}}(v_s+\phi_s+i\eta_s)\; .
\end{eqnarray}
When $v_{s}$ is non-zero, $U(1)_{Y'}$ symmetry is broken and 
$\phi_s$ becomes a massive CP-even scalar with mass-squared 
$m_{\phi}^{2}={m_{S}^{0}}^{2}+3\lambda^{0}_{S} v_{s}^{2}$. $\eta_s$, on 
the other hand, becomes the pseudoscalar Goldstone boson to be swallowed 
by the gauge boson of $U(1)_{Y'}$ , $Z^{\prime}$, to acquire a mass. 
Inserting the parametrization of $S$ in (5) into the scalar potential (2), 
one reads off the Feynman rules relating the three Higgs fields:
\begin{eqnarray}
H_{i}H_{i}\phi_{s}\; \; &:& \; \; \lambda^{0}_{iS}\, v_{s}\; \; \;\;\;
(i=1, 2)\nonumber\\
H_{i}^{c}H_{j}\phi_{s}\; \; &:& \; \; -\frac{h_{s}A_{s}^{0}}{\sqrt{2}}\; \;
\;\;\;\; (i\neq j=1,2)
\end{eqnarray}
where $H_{i}^{c}=i\sigma_{2}H_{i}^{*}$. With the help of these Feynman 
rules one obtains the tree level diagrams representing the scattering 
processes $H_{i} H_{j}\rightarrow H_{k} H_{l}$ (i, j, k, l = 1, 2) 
mediated by $\phi_{s}$. We shall not reproduce these diagrams here as 
they are similar to ones given in \cite{ma-group}. For external momenta much 
smaller than $m_{\phi}$, the scattering amplitudes mentioned above 
merely result in modification of the original quartic couplings of the Higgs 
doublets producing the effective 2HDM quartic parameters. In addition to 
these quartic coupings, the mass-squareds of the Higgs doublets do also 
change as can be seen throough the replacement of the parametrization in (5) 
into the quadratic terms in the potential (2). Defining, $\Phi_{2}=H_{2}$ 
and $\Phi_{1}=H_{1}^{c}$, the 2HDM potential below $v_{s}$ reads as follows:
\begin{eqnarray}
V(\Phi_1, \Phi_2)&=&m_{1}^{2}\Phi_{1}^{\dagger}\Phi_1+
m_{2}^{2}\Phi_{2}^{\dagger}\Phi_2+
m_{3}^{2}(\Phi_{1}^{\dagger}\Phi_2+\Phi_{2}^{\dagger}\Phi_1)+
\frac{\lambda_{1}}{2}(\Phi_{1}^{\dagger}\Phi_1)^{2}\nonumber\\&+&
\frac{\lambda_{2}}{2}(\Phi_{2}^{\dagger}\Phi_2)^{2}+
\lambda_{3}(\Phi_{1}^{\dagger}\Phi_1)(\Phi_{2}^{\dagger}\Phi_2)+
\lambda_{4}\mid \Phi_{1}^{\dagger}\Phi_2\mid ^{2}\\&+&
\frac{\lambda_{5}}{2}((\Phi_{1}^{\dagger}\Phi_2)^{2}+
(\Phi_{2}^{\dagger}\Phi_1)^{2})\nonumber
\end{eqnarray}
where the new parameters here can be expressed in terms of 
those of the underlying $Z^{\prime}$ model as follows:
\begin{eqnarray}
m_{1}^{2}&=&{m_{1}^{0}}^{2}+\frac{1}{2}\lambda^{0}_{1S}v_{s}^{2}\nonumber\\
m_{2}^{2}&=&{m_{2}^{0}}^{2}+\frac{1}{2}\lambda^{0}_{2S}v_{s}^{2}\nonumber\\
m_{3}^{2}&=&\frac{h_{s}A_{s}^{0}}{\sqrt{2}}v_{s}\nonumber\\
\lambda_{1}&=&2\lambda_{1}^{0}
-4{\lambda^{0}_{1S}}^2\frac{v_{s}^{2}}{m_{\phi}^{2}}\nonumber\\
\lambda_{2}&=&2\lambda_{2}^{0}
-4{\lambda^{0}_{2S}}^{2}\frac{v_{s}^{2}}{m_{\phi}^{2}}\\
\lambda_{3}&=&\lambda_{12}^{0}
-2\lambda^{0}_{1S}\lambda^{0}_{2S}
\frac{v_{s}^{2}}{m_{\phi}^{2}}\nonumber\\
\lambda_{4}&=&\tilde{\lambda}^{0}_{12}
-2(\frac{m_{3}^{2}}
{v_{s}})^{2}m_{\phi}^{-2}\nonumber\\
\lambda_{5}&=&-4(\frac{m_{3}^{2}}
{v_{s}})^{2}m_{\phi}^{-2}\nonumber
\end{eqnarray}
In obtaining these parameters we have used only tree level diagrams, a 
more sophisticated derivation of which would include at least the 
one-loop effects contributed by the suprsymmetric particle spectrum of 
the $Z^{\prime}$ model. However, we neglect all such loop 
contributions by assuming that these tree level results give the most 
important part of the information we need. In fact, for large enough 
soft masses (above TeV scale) one expects SUSY spectrum to have 
negligable effects at the weak scale.
 
The reduction of the scalar potential (2) to the 2HDM potential rests on
the assumption of an ordered breaking of the gauge symmetry, that is, 
only $U(1)_{Y'}$ symmetry is broken above the TeV scale while the    
SM-gauge gauge symmetry survives down to the weak scale at which it is 
broken in the usual way to reproduce the phenomenologically 
well-established electroweak data. This two-stage breaking of the gauge 
symmetry necessiates $v_{s}$, the SUSY breaking scale, be given by 
\begin{eqnarray}
v_{s}=\sqrt{-{m_{S}^{0}}^{2}/\lambda_{s}}
\end{eqnarray}
while doublets still have vanishing VEV's. With this expression for $v_s$, 
the Higgs boson $\phi_s$ and the $Z^{\prime}$ boson can be shown to 
have identical masses 
\begin{eqnarray}
m_{\phi}=M_{Z'}=\sqrt{-2{m_{S}^{0}}^{2}}
\end{eqnarray}
which has a meaning only when ${m_{S}^{0}}^{2}\, <\, 0$. Obtaining the 
pattern of VEV's $v_1=0=v_2$ and $v_{s}\neq 0$ requires a certain 
hierarchy between ${m_{S}^{0}}^{2}$ and the other mass parameters 
pertaining the Higgs doublets, that is, doublet soft mass-squareds and 
the Higgs trilinear coupling. We now turn to a detailed discussion of 
the constraints imposed on the parameters of the potential coming from 
the two-stage breaking of the gauge symmetry.
 
Our aim is to set limits on the doublet soft mass-squareds and the Higgs 
trilinear coupling for a given ${m_{S}^{0}}^{2}$ such that SM gauge 
symmetry remains unbroken at the TeV scale. For the sake of clarity it 
seems convenient to discuss the effects of these two mass scales 
seperately. Hence, we first turn off $A_{s}^{0}$ and investigate the 
ranges of ${m_{1}^{0}}^{2}$ and ${m_{2}^{0}}^{2}$. Next we shall turn to 
the discussion of $A_{s}^{0}$. 
\subsection{Effects of the soft masses}
After setting $A_{s}^{0}=0$, all of the VEV's can be solved analytically 
by requiring the potential (2) to have vanishing gradients in all 
directions in the Higgs background. In addition to the 
solution $v_1=v_2=v_s=0$ characterizing the symmetric phase, we have for 
the broken phase 
\begin{eqnarray} 
v_{1}^{2}&=&C_{0}\{ 
({\lambda^{0}_{2S}}^{2}-4\lambda^{0}_{2}\lambda^{0}_{S}){m_{1}^{0}}^{2}+
(2\lambda^{0}_{12}\lambda^{0}_{S}-\lambda^{0}_{1S}\lambda^{0}_{2S}) 
{m_{2}^{0}}^{2}\nonumber\\&+&(2\lambda^{0}_{12}\lambda^{0}_{2}- 
\lambda^{0}_{12}\lambda^{0}_{2S}){m_{S}^{0}}^{2}\}\nonumber\\
v_{2}^{2}&=&C_{0}\{
({\lambda^{0}_{1S}}^{2}-4\lambda^{0}_{1}\lambda^{0}_{S}){m_{2}^{0}}^{2}+
(2\lambda^{0}_{12}\lambda^{0}_{S}-\lambda^{0}_{1S}\lambda^{0}_{2S})
{m_{1}^{0}}^{2}\nonumber\\&+&(2\lambda^{0}_{12}\lambda^{0}_{1}-
\lambda^{0}_{12}\lambda^{0}_{1S}){m_{S}^{0}}^{2}\}\\
v_{s}^{2}&=&C_{0}\{
(2\lambda^{0}_{1S}\lambda^{0}_{2}-\lambda^{0}_{12}\lambda^{0}_{2S}) 
{m_{1}^{0}}^{2}+   
(2\lambda^{0}_{1}\lambda^{0}_{2S}-\lambda^{0}_{12}\lambda^{0}_{1S})
{m_{2}^{0}}^{2}\nonumber\\&+&({\lambda^{0}_{12}}^{2}- 
4\lambda^{0}_{1}\lambda^{0}_{2}){m_{S}^{0}}^{2}\}\nonumber
\end{eqnarray} 
where 
$1/C_{0}=4\lambda^{0}_{1}\lambda^{0}_{2}\lambda^{0}_{S}+ 
\lambda^{0}_{12}\lambda^{0}_{1S}\lambda^{0}_{2S}- 
{\lambda^{0}_{1S}}^{2}\lambda^{0}_{2}- 
{\lambda^{0}_{2S}}^{2}\lambda^{0}_{1}
-{\lambda^{0}_{12}}^{2}\lambda^{0}_{S}$ is a common factor for all three VEV's. 
From (9) it is easy to find the critical values of ${m_{1}^{0}}^{2}$ and 
${m_{2}^{0}}^{2}$ at which $v_{1}^{2}$ and $v_{2}^{2}$ vanishes:
\begin{eqnarray}
m_{1,\;  crit}^{2} &=& \frac{\lambda_{1S}^{0}}{2 \lambda_{S}^{0}} 
{m_{S}^{0}}^{2}\nonumber\\
m_{2,\;  crit}^{2} &=& \frac{\lambda_{2S}^{0}}{2 \lambda_{S}^{0}}
{m_{S}^{0}}^{2}\; .
\end{eqnarray}
Since $v_{1,2}^{2}$ change sign at these critical points they seperate 
two kinds of minima; while in one of which all VEV's in (11) are 
non-vanishing (that is, symmetry is completely broken) and in the other 
one only $U(1)_{Y'}$ is broken (that is, $v_1=0=v_2$ and $v_{s}\neq 0$). 
Since ${m_{S}^{0}}^{2}/\lambda_{s}^{0}$ must be negative for (10) be 
meaningful, the sign of the critical masses in (12) depends on the signs 
of $\lambda_{iS}^{0}$ defined in (3). This requires the knowledge of 
$h_{S}$, $g_{Y'}$ and the $U(1)_{Y'}$ charges, which would be possible 
only in a specific string or SUSY-GUT model. In addition to this 
observation, one expects Higgs doublets to have vanishing VEV's when the 
absolute values of the doublet soft mass-squareds become much smaller 
than $-{m_{S}^{0}}^{2}$. This last observation enables one to impose 
certain conditions on the Higgs VEV's (11), namely, the coefficients of 
${m_{S}^{0}}^{2}$ must be $negative$ for $v_{s}^{2}$, and $positive$ for 
$v_{1}^{2}$ and $v_{2}^{2}$ so that, for large enough $-{m_{S}^{0}}^{2}$, 
doublet VEV's obtained from (11) will be imaginary. Using 
the explicit expressions of $\lambda_{i}^{0}$ in (3), coefficients of 
${m_{S}^{0}}^{2}$ in (11) can be shown to depend on three free 
parameters; $g_{Y'}|Q'_1|/g_{Y}$, $h_{s}/G$ and $Q'_{2}/Q'_{1}$. The 
requirements about the coefficients of ${m_{S}^{0}}^{2}$ in (11) produce a 
certain constraint on these free parameters. In fact, in Fig. 1, depicted 
is the allowed region in the $g_{Y'}/g_{Y}\, - \, h_{s}/G$ plane for 
which the coefficients of ${m_{S}^{0}}^{2}$ in (11) take the required 
values mentioned above. When drawing this graph $|Q'_1|$ is absorbed 
into $g_{Y'}$, and $Q'_{2}/Q'_{1}$ is set to unity. One recalls that,for 
$Q'_{2}/Q'_{1}=1$, $\lambda_{1}^{0}=\lambda_{2}^{0}$ and 
$\lambda_{1S}^{0}=\lambda_{2S}^{0}$ as is seen from (3). Although 
we consider a specific choice for $Q'_{2}/Q'_{1}$ here, with general 
formulae for VEV's in (11), one can investigate other possibilities 
for $Q'_{2}/Q'_{1}$ as well. In forming Fig.1 we let $g_{Y'}/g_{Y}$ and 
$h_{s}/G$ vary from 0.5 to 1.5 which is a symmetric interval around 
unity. We see that in this interval allowed $h_{s}/G$ values      
increase in approximate proportion with the allowed  $g_{Y'}/g_{Y}$ 
values.   

Now we turn to the main question of how large doublet soft 
mass-squareds, in comparison with ${m_{S}^{0}}^{2}$, can be to have 
SM gauge symmtery unbroken at the TeV scale. We illustrate the ranges of 
the doublet soft mass-squareds with the help of the information provided by 
Fig. 1. Out of all the candidate points shown in Fig. 1 we take the one 
for which $h_{s}/G=1$ and $g_{Y'}/g_{Y}=1$. This choice is in agreement
with the usual prescriptions about the low-energy $Z^{\prime}$ models, that 
is $g_{Y'}\sim g_{Y}$ \cite{haber-sher} and $h_{s}\sim 0.7$ \cite{lang}. 
As metioned before, for $Q'_{2}/Q'_{1}=1$ used in forming Fig. 1, 
$\lambda_{1S}^{0}=\lambda_{2S}^{0}$ for which the critical values of the 
doublet soft mass-squareds in (12) are equal. As the critical points are 
identical, for simplicity of the illustration, we take 
${m_{1}^{0}}^{2}={m_{2}^{0}}^{2}=m_{0}^{2}$ and show the  
$m_{0}^{2}/|{m_{S}^{0}}^{2}|$ dependence of the squared VEV's (11) in 
units of $|{m_{S}^{0}}^{2}|$, in Fig. 2. As is seen from this figure, 
$v_{s}^{2}$ (solid line) increases monotonically with increasing
$m_{0}^{2}/|{m_{S}^{0}}^{2}|$, never going through zero. $v_{1}^{2}$ 
and $v_{2}^{2}$ (dashed line), however, remain identical and drop below 
zero at approximately $m_{0}^{2}/|{m_{S}^{0}}^{2}|=-0.58$ as exactly 
predicted by the critical points in (12). Thus, Fig. 2 can be concluded 
by saying that SM gauge symmetry remains unbroken as long as 
$m_{0}^{2}/|{m_{S}^{0}}^{2}|>-0.58$ provided the parameter set used 
for Fig. 1 and ${m_{1}^{0}}^{2}={m_{2}^{0}}^{2}$ are assumed. 

The more general case of $Q'_{2}/Q'_{1}\neq 1$ and 
${m_{1}^{0}}^{2}\neq {m_{2}^{0}}^{2}$ can be analyzed in a similar way by 
using the expressions of VEV's in (11), and the critical points in (12). 
In this case there will be more free variables ($v_1\neq v_2$) and their  
behaviour will be complicated for graphical presentation. Despite this, 
when the coefficient of ${m_{S}^{0}}^{2}$ is $positive$ for doublet VEV's 
and $negative$ for SM -singlet VEV, and when the doublet mass-squareds 
exceed their respective critical points given in (12), one has a minimum of 
the potential for which SM gaue symmetry remains unbroken at the scale 
$v_{s}\sim \sqrt{{m_{S}^{0}}^{2}}\sim {\cal{O}}$(TeV). This last 
statement summarizes the results about the minimum of the potential 
leading to a 2HDM at the weak scale when the Higgs trilinear coupling 
mass parameter $A_{s}^{0}$ is much smaller than the quadratic mass 
parameters of the Higgs doublets.

\subsection{Effects of the Higgs trilinear coupling} 
The basic property of the Higgs trilinear coupling $A_{s}^{0}$ is that it 
forces all fields $H_1$, $H_2$, and $S$ to have identical VEV's, 
depending on its strength compared to the other mass parameters in the 
potential \cite{lang}. This happens especially when all mass-squareds are 
of the same order and much smaller than ${A_{s}^{0}}^{2}$, and type of 
the transition from the symmetric to the broken minimum depends the sign 
of the sum $m^{2}={m_{1}^{0}}^{2}+{m_{2}^{0}}^{2}+{m_{S}^{0}}^{2}$. The 
critical point at which the transition occurs is 
$A_{s}^{crit}=(8/3)m^{2}$, and for positive (negative) $m^{2}$ 
passage to the broken mininum is a first (second) order phase transition 
\cite{dur}. In the present case, where $-{m_{S}^{0}}^{2}$ is much larger 
than all other mass parameters in the potential (2), one is to analyze 
the doublet VEV's $v_{1,2}$ for a given $v_{s}$ to find the critical 
value of $A_{s}^{0}$ at which the transition from SM- symmetric minimum 
to SM-broken minimum occurs. This requires the solution of the 
minimization equations for the potential (2) in $v_{1,2}$ direction. Here 
one faces with two coupled third order algebraic equations the analytic 
solution of which is hard to construct, and even if constructed, the 
results will not be transparent. Thus, we consider now a special but 
an important limit, that is, $v_1=v_2$, which can easily be obtained for 
certain set of parameters as illustrated in the last section. In this case 
it is not hard to determine the critical point $A_{s}^{crit}$ at which 
the transtion to an SM- broken phase occurs: 
\begin{eqnarray}
A_{s}^{crit}=\frac{1}{\sqrt{2}h_{S}v_{s}}({m_{1}^{0}}^{2}+ 
{m_{2}^{0}}^{2}+  
\frac{(\lambda^{0}_{1S}+\lambda^{0}_{2S})}{2} v_{s}^{2})
\end{eqnarray}
where $v_{s}$ is given by (9). An inspection on this equation reveals 
that $A_{s}^{crit}$ is large (small) when ${m_{1}^{0}}^{2}+{m_{2}^{0}}^{2}$
is positive (negative). Moreover, when 
$|{m_{1}^{0}}^{2}+{m_{2}^{0}}^{2}|$ is small $A_{s}^{crit}$ is mainly 
determined by  $v_{s}$. Moreover, the construction of $A_{s}^{crit}$ from 
the minimization conditions reveals that transition is always second order. 
We now illustrate the range of  $A_{s}^{0}$ for two distinct choices for 
the doublet soft mass-squareds. In Fig. 3, depicted are the $A_{s}^{0}/m_{0}$ 
dependence of $v_{s}/m_{0}$ (solid curve) and $v_{1,2}/m_{0}$ (dashed 
curve) for  ${m_{S}^{0}}^{2}= -25 m_{0}^{2}$, and other parameters are 
the same as ones used for Fig. 2. Here the choice of ${m_{S}^{0}}^{2}$ is 
arbitrary and sufficient as long as it is large negative compared to other 
parameters of mass dimension. As is seen from Fig. 3, the Higgs doublets 
have vanishing VEV's until $A_{s}^{0}$ reaches the transition point 
$A_{s}^{crit}=3 m_{0}$. Meanwhile $v_{s}$ remains to have the value given by 
(9). The critical point is exactly the one predicted by (13). To see the 
dependence of $A_{s}^{crit}$ in (13) on the doublet mass parameters we 
now illustrate the case of ${m_{S}^{0}}^{2}= -25 m_{0}^{2}$ and 
${m_{1}^{0}}^{2}={m_{2}^{0}}^{2}=25m_{0}^{2}$ with other parameters 
being as in Fig. 1. In this case, as is seen from Fig. 4, critical point 
is large, $A_{s}^{crit}=7.65 m_{0}$, and situated at the point predicted 
by (13). 

In the light of the above observations, we conclude that when the Higgs 
trilinear coupling $A_{s}^{0}$ is below a certain critical value 
$A_{s}^{crit}$ the Higgs doublets have vanishing VEV's. This critical 
point depends on the doublet soft mass-squareds and $v_{s}$. When the 
parameters of the potential allow for $v_1=v_2$, the critical value of 
$A_{s}^{0}$ has the expression given in (13). It should be noted that, 
when ${m_{1}^{0}}^{2}$ and ${m_{2}^{0}}^{2}$ are positive, the Higgs 
doublets are unable to develop VEV's with the help of their masses, 
leaving $A_{s}^{0}$ as the only remedy. This requires $A_{s}^{0}$ to be 
large enough ($\sim \sqrt{{m_{S}^{0}}^{2}}$) to help all fields acquire 
asymptotically nearly equal VEV's as is clearly seen from Fig. 3 and Fig. 4. 
\section{Properties of the resulting 2HDM}
In the last section we obtained an effective 2HDM potential after 
breaking the extra $U(1)$ at the TeV scale. Equation (7) represents the 
most general 2HDM potential with the parameters in (8). We now discuss 
some important prioperties of (7).
\begin{itemize}
\item There are three parameters of mass dimension in (7): $m_{1}^{2}$, 
$m_{2}^{2}$ and $m_{3}^{2}$ each of which has its own characteristic 
minimum. As it is convenient to discuss their effects seperately, let us 
first consider the case of $|m_{3}^{2}| << |m_{1,2}^{2}|$. Needless to 
say this case corresponds to the small trilinear coupling regime of the 
potential (2), discussed in Sec. 2.1. In this case 
the neutral componenet of the Higgs doublet $\Phi_i$ develops a VEV 
$<\Phi_{i}^{0}>\sim \sqrt{-m_{i}^{2}/\lambda_{i}}$ where the mixed quartic 
terms are neglected. Clearly, if $m_{i}^{2} > 0$, VEV becomes imaginary 
and potential prefers the symmetric minimum; $<\Phi_{i}^{0}>= 0$. There 
are two ways of making $m_{i}^{2}$ negative. First, the 
soft mass ${m_{1}^{0}}^{2}$ can be negative. Second $\lambda^{0}_{iS}$
can be negative depending on the relative magnitudes of its D-term and 
F-term components, as listed in (3). Fig. 2, which is obtained for a 
special set of parameters, shows a typical case applicable when  
$|m_{3}^{2}|$ is small. When soft masses of the Higgs doublets are negative, 
as is necessary to have $<\Phi_{i}^{0}>\neq 0$, one can vary 
$m_{1,2}^{2}$ up to $\sim 0.5 v_{s}^{2}$, in accordance with (12) and 
Fig. 2.	On the other hand, the case of negative $\lambda^{0}_{iS}$ can be 
obtained when $-g_{Y}^{2}Q'_{i}Q_{S}$ is larger than $h_{S}^{2}$. This 
latter alterative should be taken with care because perturbative nature 
of the model at the TeV scale can be disturbed. Of course, having at 
least one of $m_{1}^{2}$ and $m_{2}^{2}$ is sufficient, thanks to the 
mixed quartic couplings in (7), one can make both VEV's non zero. If 
$|m_{1}^{2}| << -m_{2}^{2}$ one gets a large $\tan\beta$ minimum because 
$<\Phi_{1}^{0}>$ will be much smaller than $<\Phi_{2}^{0}>$. In the 
opposite case, $|m_{2}^{2}| << -m_{1}^{2}$ one has a small $\tan\beta$ by 
the similar arguments. While in former case the huge $m_{t}/m_{b}$ ratio 
can be obtained for nearly equal top ($h_{t}$) and bottom ($h_{b}$) Yukawa 
couplings, in the latter case one would need $h_{t} << h_{b}$. Finally, 
$m_{1}^{2} \sim m_{2}^{2}$ implies $\tan\beta\sim 1$ when both 
mass-squareds are negative, and in this case $h_{t}\sim 1 >> h_{b}$ is 
sufficient. 
 
When $|m_{3}^{2}| >> |m_{1,2}^{2}|$, $m_{3}^{2}$ makes up almost a unique 
mass scale for the potential. Then $<\Phi_{1}^{0}>\sim <\Phi_{2}^{0}>$ 
and consequently $\tan\beta\sim 1$ for which $h_{t}\sim 1 >> h_{b}$ as 
mentioned above. This large $|m_{3}^{2}|$ case correspond to the large 
$A_{s}^{0}$ regime analyzed in Sec. 2.2.

These naive tree level considerations are not valid when the loop effects on 
the parameters of the potential (7) are taken into account. The largest of 
such corrections come to the quartic coupling of $H_{2}$, $\lambda_{2}$ 
due to large $h_{t}$ \cite{dur}. This change in $\lambda_{2}$ directly 
affects $v_{2}$ wherby modifying the value of $\tan\beta$. However, in 
the large $|m_{3}^{2}|$ regime one does not expect too big changes, as 
long as it still dominates over the quadratic mass parameters. 

\item Depending on the way $U(1)_{Y'}$ is broken, the various operators 
in the 2HDM potential (7) get suppressed or amplified, as are seen from (8). 
When $A_{s}^{0}$ is vanishingly small, as discussed in Sec. 2.1, 
$\lambda_5$ becomes negligable and $\lambda_4$ approaches  
$\tilde{\lambda}_{12}^{0}$. In addition, the quadractic mass parameters 
$m_{1,2}^{2}$ approach to the corresponding soft mass-squareds 
in (2). Finally, the Higgs mixing mass $m_{3}^{2}$ becomes negligably 
small in this limit. 
  
When $A_{s}^{0}$ is the dominant mass parameter, as analyzed in Sec. 2.2, 
one has $m_{3}^{2}\sim m_{1}^{2} \sim m_{2}^{2} \sim {A_{s}^{0}}^{2}$. 
Unlike the previous case, here  $\lambda_5$ and $m_{3}^{2}$ are no way 
negligable.

Finally, it is worthy of noting that the values of $\lambda_{1}$, 
$\lambda_{2}$ and $\lambda_{3}$ is independent of the way $U(1)_{Y'}$ 
symmetry is broken. This is because 
$v_{s}^{2}/m_{\phi}^{2}= 1/2\lambda_{S}^{0}$ in both cases.
\item Throughout the discussions in the last section all  
parameters of the potential (2) have been assumed real. In addition to 
this, the Higgs fields were assigned real eigenvalues. If the 
Higgs trilinear coupling $A_{s}^{0}$ in the potential (2) were complex so 
are $\lambda_{5}$ and $m_{3}^{2}$ in (7). After writing the 
relevant parts of the potential (7) appropriately, one observes that 
under a CP transformation $\Phi_{i}\rightarrow 
e^{i\theta_{i}}\Phi_{i}^{*}$, $\lambda_{5}$ and $m_{3}^{2}$ dependent 
terms do not transform trivially. However, 
since $\lambda_{5}/\lambda_{5}^{*} = (m_{3}^{2}/{m_{3}^{2}}^{*})^{2}$, 
CP- invariance of the potential is guaranteed \cite{froggat2}. In fact, 
there are strong experimental limits, though in MSSM, forcing $m_{3}^{2}$ 
to be nearly real \cite{reality}. Thus, the resulting 2HDM potential (7) 
does not support the explicit violation of CP. As is seen from (8), 
$\lambda_{5}< 0$, so that the 2HDM potential (7) cannot accomodate 
spontaneous CP violation too \cite{cp}. Thus, the resulting 2HDM 
does break CP neither explicitely nor spontaneously.

\item As has already been discussed in \cite{lang,dur}, the parameters of 
the $Z^{\prime}$ model (2) can be connected to the unification level 
(strings or GUT's) initial conditions via the RGE's. In this sense, better 
the determinations at low-energies, better the knowledge one has about 
unification level parameters. Thus, it would be desirable to constrain 
the parameters of the 2HDM potential (7) using the low energy data as 
much as possible. For example, the precisely measured Z-pole observables 
can be used to constrain these parameters \cite{froggat2,froggat3}. 
\end{itemize}
\section{Conclusions}
In this work, we have presented the reduction of a supersymmetric 
TeV-scale extra $U(1)$ to an effective 2HDM at the weak scale. We have 
identified the appropriate portions of the parameter space of the extra 
$U(1)$ model for obtaining a 2HDM below the TeV scale. Moreover, 
properties of the resulting 2HDM is connected to the mechanism of extra 
$U(1)$ breaking. In addition to these, we have discussed various 
properties of this 2HDM in connection with weak scale phenomenology, $Z$- 
pole data and CP -violation. Other than these $Z^{\prime}$ models, 
reduction of NMSSM to an effective 2HDM can also be worked out. In spite 
of the cosmological problems due to the broken $Z_3$ symmetry 
\cite{zeldovic}, from the particle physics point of view such models can 
provide us with an effective 2HDM at the weak scale, when the SM - 
singlet picks up a VEV around a TeV. In this case, mass of the 
pseudoscalar boson breaking the unwanted Peccei-Quinn symmetry will be an 
important constraint. Whatever the TeV scale model we start with, the 
weak scale 2HDM can constrain the parameter space of the original model 
through the experimental data. This low-energy determination can 
eventually be useful in constraining the unification scale initial 
conditions.

Author would like to thank to referee for his/her constructive remarks. 

\newpage
\begin{center}
{\bf Figure Captions}
\end{center}
Fig. 1. The region in the $g_{Y'}/g_{Y}\, - \, h_{s}/G$ plane for which 
the coefficient of ${m_{S}^{0}}^{2}$ in (9) is negative for SM-singlet VEV 
and positive for the doublet VEV's.\\
Fig. 2. $m_{0}^{2}/|{m_{S}^{0}}^{2}|$ dependence of the squared VEV's (9)
in units of $|{m_{S}^{0}}^{2}|$ for ${m_{1}^{0}}^{2}= 
{m_{2}^{0}}^{2}=m_{0}^{2}$ and $h_{s}/G=1$ and $g_{Y'}/g_{Y}=1$. Here 
solid curve is for $v_{s}$, and dashed curve fo $v_{1,2}$.\\
Fig. 3. $A_{s}^{0}/m_{0}$ dependence of $v_{s}$ (solid curve) and 
$v_{1,2}$ (dashed curve). Here ${m_{S}^{0}}^{2}= -25 m_{0}^{2}$ and 
other parameters are as in Fig. 2.\\
Fig. 4. $A_{s}^{0}/m_{0}$ dependence of $v_{s}$ (solid curve) and
$v_{1,2}$ (dashed curve). Here ${m_{S}^{0}}^{2}= -25 m_{0}^{2}$, 
${m_{1}^{0}}^{2}={m_{2}^{0}}^{2}=25m_{0}^{2}$, and other parameters are 
as in Fig. 1.
\newpage
\begin{figure}
\vspace{10cm}
\end{figure}
\begin{figure}
\vspace{12.0cm}
    \includegraphics{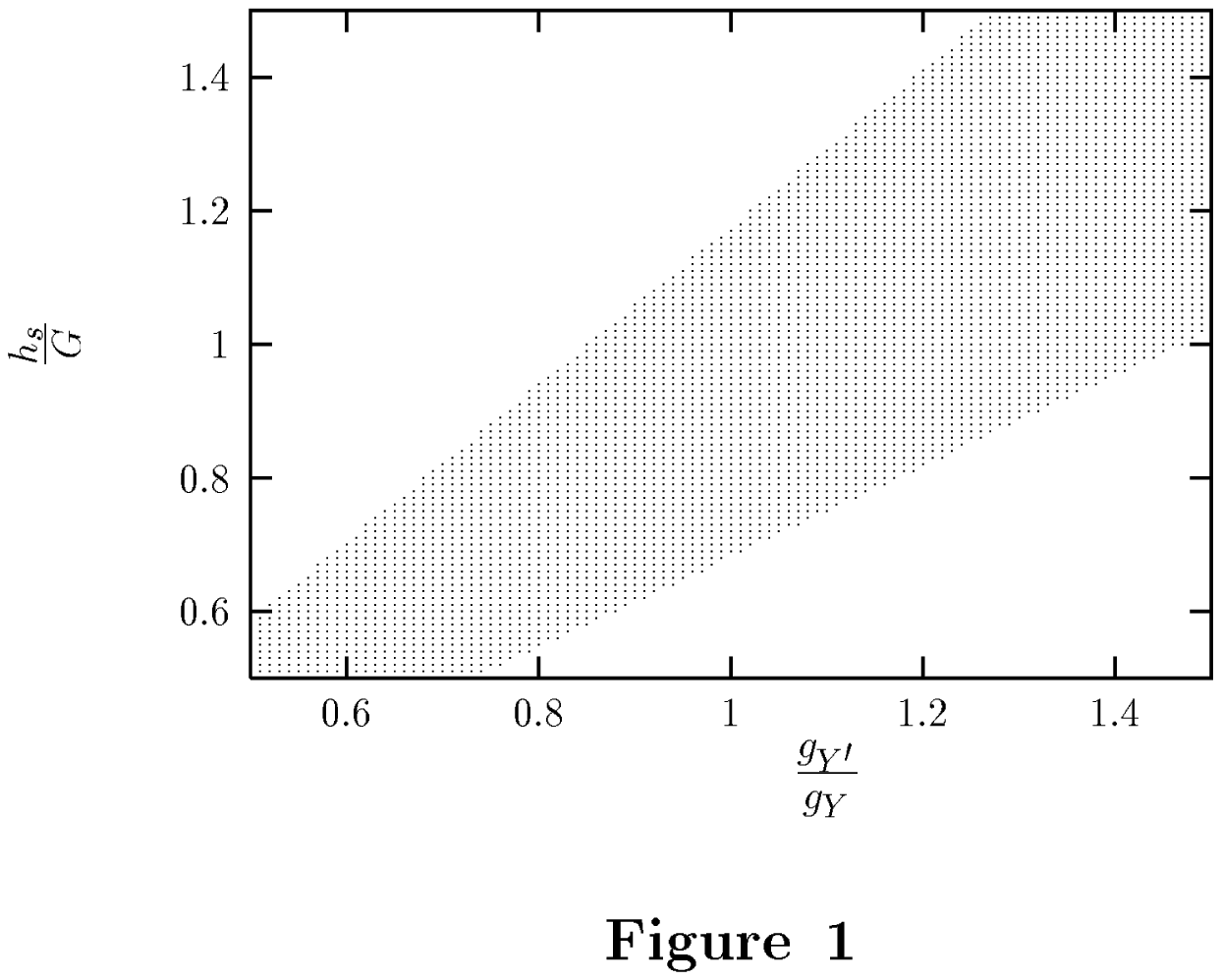}
    \vspace{-8.0cm}
\vspace{0.0cm}
\end{figure}
\begin{figure}
\vspace{10.0cm}
\end{figure}
\begin{figure}
\vspace{12.0cm}
    \includegraphics{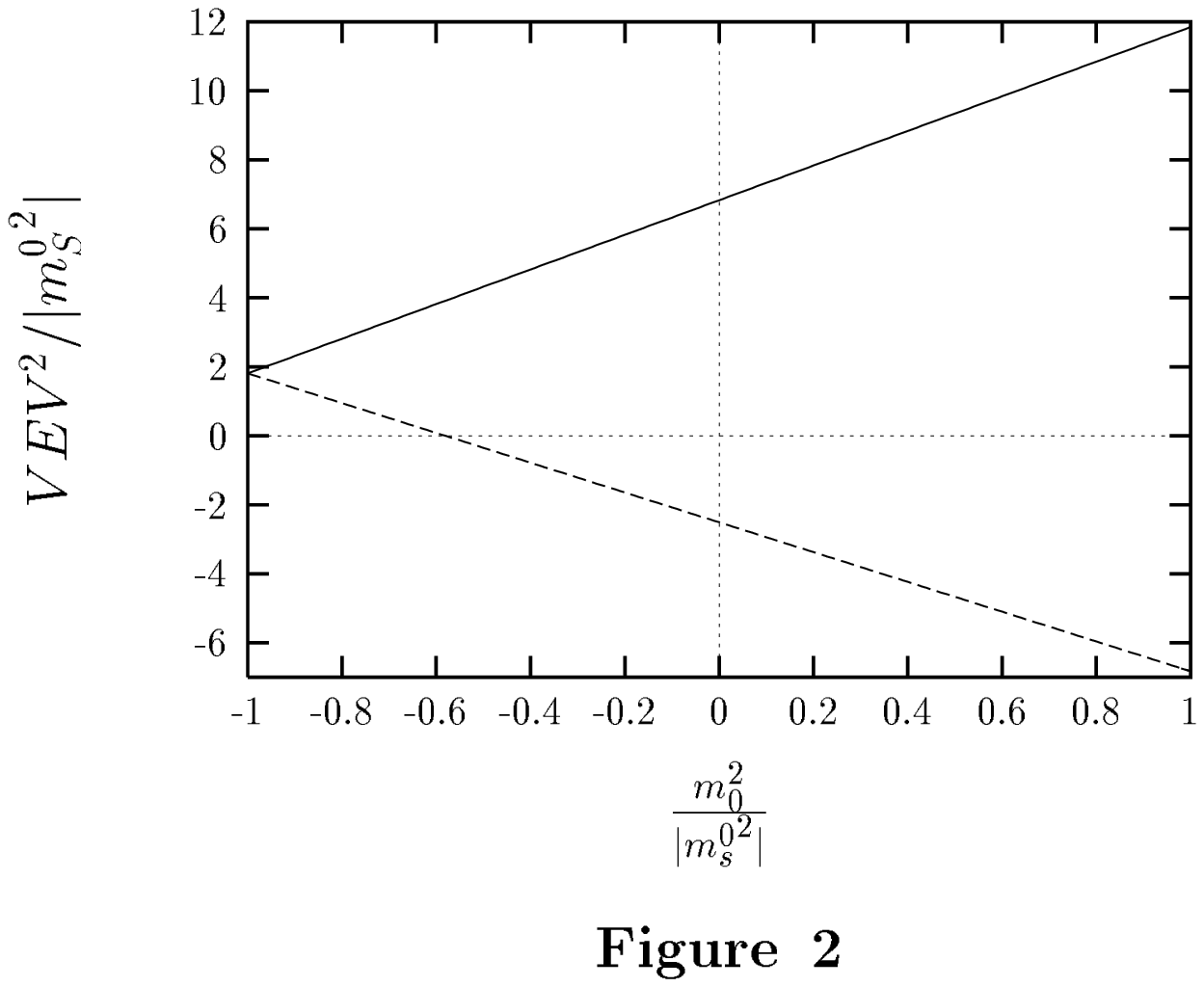}
    \vspace{-8.0cm}
\vspace{0.0cm}
\end{figure}
\begin{figure}
\vspace{10.0cm}
\end{figure}
\begin{figure}
\vspace{12.0cm}
    \includegraphics{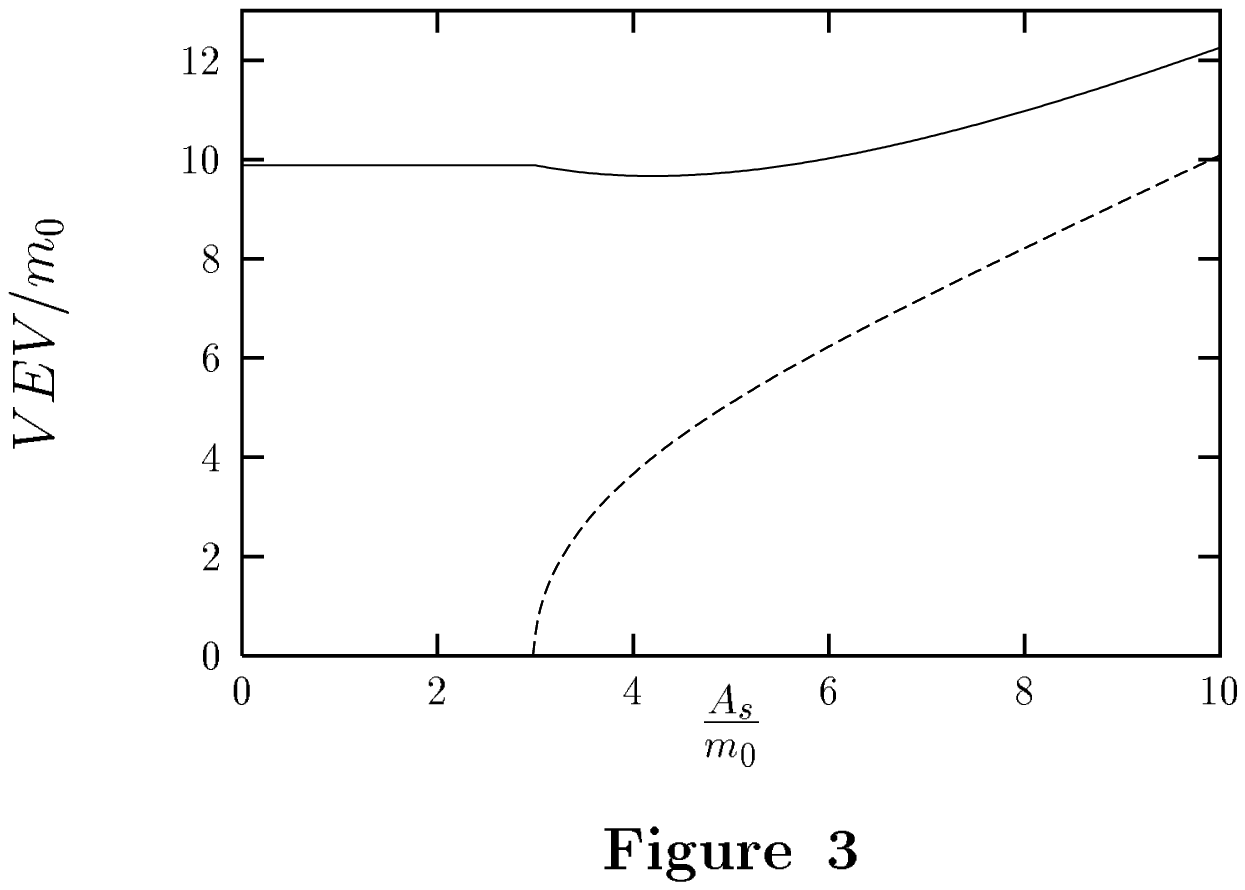}
    \vspace{-8.0cm}
\vspace{0.0cm}
\end{figure}
\begin{figure}
\vspace{10.0cm} 
\end{figure}
\begin{figure}
\vspace{12.0cm}
    \includegraphics{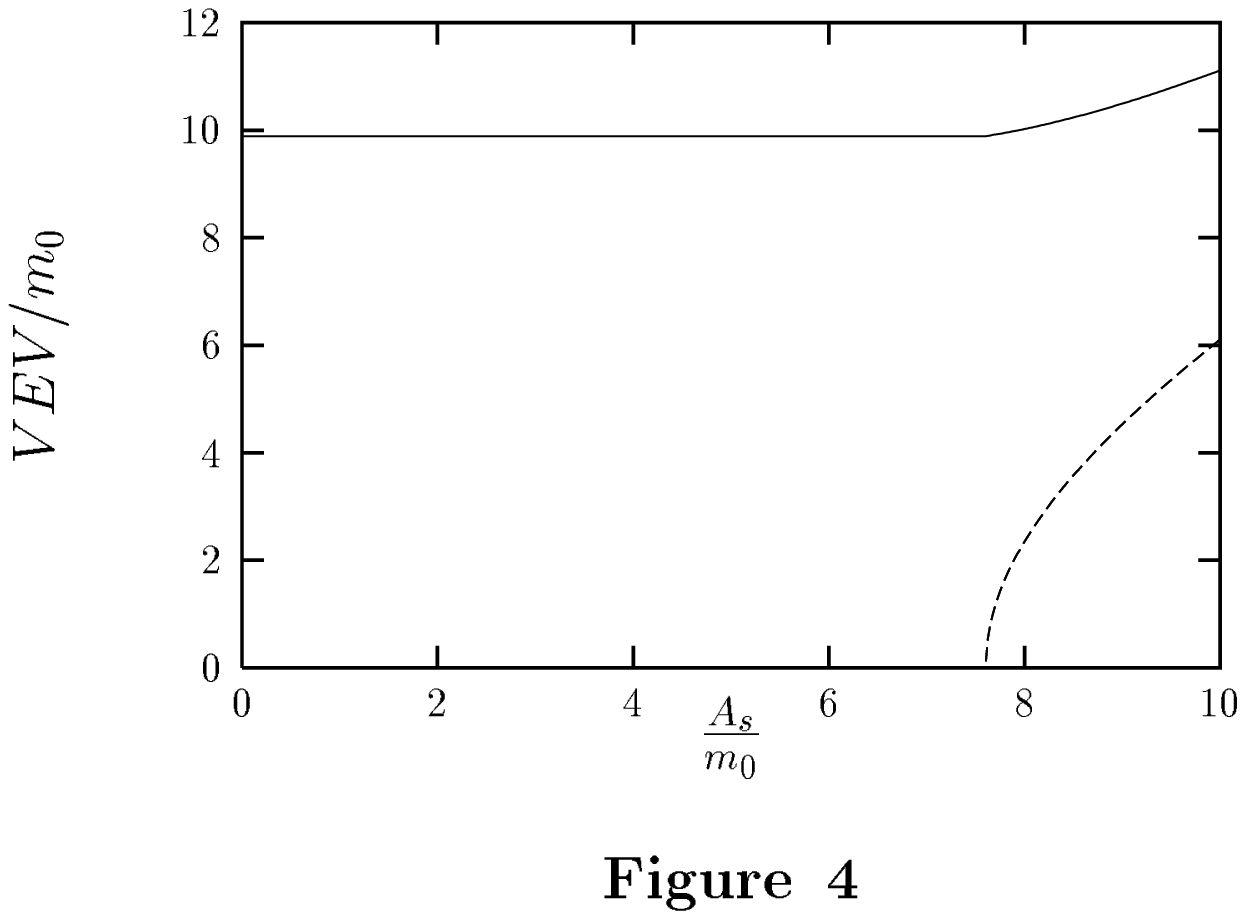}
    \vspace{-8.0cm}
\vspace{0.0cm}
\end{figure}

\begin{thebibliography}{99}
\bibitem{haber} {J. F. Gunion, H. E. Haber, Nucl. Phys. {\bf B272} (1986) 
1.} 
\bibitem{nilles} {H. P. Nilles, Phys. Rep. {\bf 110} (1984) 1.}
\bibitem{froggat} { C. D. Froggatt, I. G. Knowles, R. G. Moorhouse, Phys. 
Lett. {\bf B249} (1990) 273.}
\bibitem{kane} {H. E. Haber, G. L. Kane, T. Sterling, Nucl. Phys. {\bf 
B161} (1979) 493.}
\bibitem{sakharov} {A. Sakharov, JETP Lett. {\bf 5} (1967) 24.}
\bibitem{kaplan}{ A. G. Cohen, D. B. Kaplan, A. E. Nelson, Phys. Lett. 
{\bf B263} (1991) 86.} 
\bibitem{suematsu} {D. Suematsu, Y. Yamagishi, Int. J. Mod. Phys. {\bf 
A10} (1995) 4521.}
\bibitem{lang} {M. Cvetic, D. A. Demir, J. R. Espinosa, L. Everett, P.
Langacker, Phys. Rev. {\bf D56} (1997) 2861.}
\bibitem{jose} {J. R. Espinosa, Nucl. Phys. Proc. Suppl. {\bf 62} (1998) 
187.}
\bibitem{dur} {D. A. Demir, N. K. Pak, Phys. Rev. {\bf D57} (1998) 6609.}
\bibitem{cvetic-lang} {M. Cvetic, P. Langacker, Phys. Rev. {\bf D54} 
(1996) 3570.} 
\bibitem{ma-group} {X. Li, E. Ma, J.Phys. {\bf G23} (1997) 885; E. Ma, D. 
Ng, Phys. Rev. {\bf D49} (1994) 6164; T. V. Duong, E. Ma, Phys. Lett. 
{\bf B316} (1993) 307; E. Keith, E. Ma, Phys. Rev. {\bf D54} (1996) 3587; 
E. Keith, E. Ma, Phys. Rev. {\bf D56} (1997) 7155.} 
\bibitem{haber-sher} {H. E. Haber, M. Sher, Phys. Rev. D35 (1987) 2206.}
\bibitem{froggat2} {C. D. Froggatt, R. G. Moorhouse, I. G. Knowles, 
Nucl. Phys. {\bf B386} (1992) 63.}
\bibitem{reality} {T. Falk, K. A. Olive, M. Srednicki, Phys. Lett. {\bf 
B354} (1995) 99; M. Dugan, B. Grinstein, L. Hall, Nucl. Phys. {\bf B255} 
(1985) 513.}
\bibitem{cp} {T. D. Lee, Phys. Rev. {\bf D8} (1972) 1226, Phys. Rep. {\bf 
9} (1974) 143; Y. L. Wu, L. Wolfenstein, Phys. Rev. Lett. {\bf 73} (1994) 
1762.}
\bibitem{froggat3} {C. D. Froggatt, R. G. Moorhouse, I. G. Knowles, Phys. 
Rev. {\bf D45} (1992) 2471.}
\bibitem{zeldovic} {Ya. B. Zel'dovich, I. Yu. Kobzarev, L. B. Okun, Sov. 
Phys. JETP {\bf 40} (1975) 1.}
\end{thebibliography}
\end{document}